# Estimation of dilution of a Fast Faraday Cup response due to the finite particles speed

A. Shemyakin**, Fermilab*, Batavia, IL 60510, USA

*Abstract*
Deviation of the Fast Faraday Cup signal from the longitudinal shape of the measured bunch is estimated, and a simple formula for the increase of the signal rms width is suggested.

## 1. Introduction: a Fast Faraday Cup

The length of a charge particle bunch can be measured from duration of a pulse generated when the bunch is intercepted by an electrode (addressed as a collector in this paper). If the bunch particles are non-relativistic, the bunch electric field causes motion of charges in the collector well before arrival of the bunch itself. As a result, for a bunch with the length short in comparison with the typical sizes of the vacuum chamber the pulse duration may be determined mainly by the chamber size. One of the ways to counteract the effect is to cut a beamlet out of the bunch by an orifice and place the beamlet collector close to the orifice. Such device is called a Fast Faraday Cup (FFC, e.g. [1]). In this case, one can expect the elongation of the pulse to be determined by a factor of $\kappa \cdot \tau$, where $\tau = d/v$ is the time of flight through FFC, $d$ is the distance between the orifice and the collector, $v$ is the particles velocity, and $\kappa$ is a numerical coefficient determined by the FFC geometry. This paper estimates this coefficient in a simple model.

## 2. Model

A model of a FFC is shown in Fig. 1. The bunch travels from the right; only a beamlet crosses the boundary of the orifice and is deposited to the collector. The current flowing from the collector to "true ground" is measured and used to characterize the bunch length.

The estimations in this paper are made under following assumptions:
1. The diameters of the orifice and a possible hole in the collector are small in comparison with the distance between the electrodes $d$. Therefore, the model assumes that electric field at the collector surface is affected only by particles between the orifice and the collector (i.e. with the longitudinal coordinate $z$  $0<z<d$).
2. The beamlet's transverse size has no effect on the image charge at the collector. Trajectories are straight lines.
3. There are no secondary particles emitted or scattered from the collector surface.
4. Particles are slow enough so the estimations can be made in electrostatic approximation. Magnetic fields and electrodes resistivity are neglected.

---

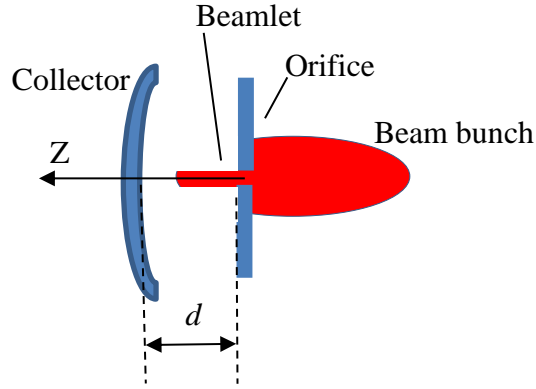

Figure 1. FFC model.

In this model, the current flowing from the collector $I_c(t)$ to ground has two components. One of them is the direct current of the beam. Let us choose $t=0$ at the moment of crossing the orifice boundary by the bunch's center, i.e.

$$\int_{-\infty}^{\infty} I_b(t) \cdot t \, dt = 0 \,, \tag{1}$$

so that the first component is $I_b(t-\tau)$. The second component is related to changes in the image charge $Q_c(t)$ at the collector surface. This charge exists all the time when there are charged particles in the space between the orifice and the collector, and it is the reason for the pulse shape dilution in FFC measurements. Together,

$$I_c(t) = I_b(t-\tau) - \dot{Q}_c(t) \,, \tag{2}$$

where the dot indicate the time derivative.

## 3. Expression for the collector signal rms width

One can consider a unity point charge at the distance of $z$ from the orifice surface along the beam trajectory. Such charge creates images charges at both electrodes, which sum is -1. Denoting $q_c(z)$ the portion of the charge at the collector surface, the total image charge created by a bunch on the collector is a sum of image charges created by all particles between two electrodes:

$$Q_c(t) = \int_0^d q_c(z) I_b(t-z/v) \frac{1}{v} dz \tag{3}$$

Correspondingly, the pulse recorded from the collector is calculated as

$$I_c(t) = I_b(t-\tau) - \int_0^d q_c(z) \dot{I}_b(t-z/v) \frac{1}{v} dz \tag{4}$$

To quantify the pulse elongation, let us compare the rms lengths of the measured pulse $\sigma_c$ and of the bunch $\sigma_b$. By definition,

$$\sigma_c^2 = \frac{1}{Q_b} \int_{-\infty}^{\infty} t_1^2 I_c(t_1) dt_1 - \left[ \frac{1}{Q_b} \int_{-\infty}^{\infty} t_1 I_c(t_1) dt_1 \right]^2 \,, \tag{5}$$

where

$$Q_b = \int_{-\infty}^{\infty} I_c(t_1) dt_1 \qquad (6)$$

is the bunch charge. For further calculations, we can use the natural assumption that $I_b(t)$ decreases fast toward both infinities and simplify expressions for momenta of $\dot{I}_b(t)$ taking into account definitions of Eq.(1) and (6):

$$\int_{-\infty}^{\infty} t_1^2 \dot{I}_b(t_1 - z/v) dt_1 = \int_{-\infty}^{\infty} (t_2 + z/v)^2 \dot{I}_b(t_2) dt_2 =$$

$$= (z/v)^2 \int_{-\infty}^{\infty} \dot{I}_b(t_2) dt_2 + 2z/v \int_{-\infty}^{\infty} t_2 \dot{I}_b(t_2) dt_2 + \int_{-\infty}^{\infty} t_2^2 \dot{I}_b(t_2) dt_2 =$$

$$= 0 - 2z/v \int_{-\infty}^{\infty} I_b(t_2) dt_2 - 2 \int_{-\infty}^{\infty} t_2 I_b(t_2) dt_2 = -2z/v \cdot Q_b; \qquad (7)$$

$$\int_{-\infty}^{\infty} t_1 \dot{I}_b(t_1 - z/v) dt_1 = \int_{-\infty}^{\infty} (t_2 + z/v) \dot{I}_b(t_2) dt_2 = (z/v) \int_{-\infty}^{\infty} \dot{I}_b(t_2) dt_2 + \int_{-\infty}^{\infty} t_2 \dot{I}_b(t_2) dt_2 = -Q_b.$$

Now we can substitute Eq.(4) into Eq.(5). Using Eq.(7), the first integral of Eq. (5) is transformed as

$$\int_{-\infty}^{\infty} t_1^2 I_c(t_1) dt_1 = \int_{-\infty}^{\infty} t_1^2 I_b(t_1 - \tau) dt_1 - \int_{-\infty}^{\infty} t_1^2 \int_0^d q_c(z) \dot{I}_b(t_1 - z/v) \frac{1}{v} dz\, dt_1 =$$

$$= \int_{-\infty}^{\infty} (t_3 + \tau)^2 I_b(t_3) dt_3 - \int_0^d q_c(z) \frac{1}{v} dz \int_{-\infty}^{\infty} t_1^2 \dot{I}_b(t_1 - z/v) dt_1 = \qquad (8)$$

$$= \int_{-\infty}^{\infty} t_3^2 I_b(t_3) dt_3 + \tau^2 Q_b + \frac{2Q_b}{v^2} \int_0^d q_c(z) z\, dz = Q_b \left[ \sigma_b^2 + \tau^2 \left( 1 + 2 \int_0^1 q_c(d \cdot u) u\, du \right) \right]$$

and the second one as

$$\int_{-\infty}^{\infty} t_1 I_c(t_1) dt_1 = \int_{-\infty}^{\infty} t_1 I_b(t_1 - \tau) dt_1 - \int_{-\infty}^{\infty} t_1 \int_0^d q_c(z) \dot{I}_b(t_1 - z/v) \frac{1}{v} dz\, dt_1 =$$

$$= \int_{-\infty}^{\infty} (t_3 + \tau) I_b(t_3) dt_1 - \int_0^d q_c(z) \frac{1}{v} dz \int_{-\infty}^{\infty} t_1 \dot{I}_b(t_1 - z/v) dt_1 = \qquad (9)$$

$$= \tau \cdot Q_b + \tau \cdot Q_b \int_0^1 q_c(d \cdot u) du$$

Combining Eq.(8) and Eq.(9), Eq.(5) is written as

$$\sigma_c^2 = \sigma_b^2 + \tau^2(1 + 2A_1) - \tau^2(1 + A_0)^2 \equiv \sigma_b^2 + \sigma_{FFC}^2 \qquad (10)$$

where for brevity the integrals are notated as

$$A_0 = \int_0^1 q_c(d \cdot u) du; \quad A_1 = \int_0^1 q_c(d \cdot u) u\, du. \qquad (11)$$

Therefore, the rms width of the collector signal is the sum, in quadrature, of the rms width of the bunch shape and the component describing the FFC response $\sigma_{FFC} \equiv \kappa \cdot \tau$ with the dimensionless geometrical factor $\kappa$ is calculated as

$$\kappa = \sqrt{2A_1 - 2A_0 - A_0^2} \tag{12}$$

The value of the $\sigma_{FFC}$ describes the rms width of a signal generated by a point charge moving between the orifice and the collector.

## 4. Calculation for simple geometries

In cases when the FFC electrodes can be approximated by a 1D geometry, the geometrical coefficient in Eq.(12) can be expressed in elementary functions.

- *Plane electrodes*

When the FFC electrodes are flat with transverse dimensions much larger than the gap $d$, it can be approximated by two parallel infinite plates. To calculate the function $q_p(z)$, let us note that portion of the image charge at the collector plane is the same for all particles located at the same $z$ independently on other two coordinates. Therefore, an infinitely thin plane charged with the surface charge density $s_0$ and located at $z$ produces the image charge density at the collector plane of $s_c = s_0 \cdot q_p(z)$ and $s_{or} = -s_0 - s_c$ at the orifice plane. Condition that the collector and orifice plane have the same potential can be written as

$$(s_{or} - s_c - s_0) \cdot z + (s_{or} - s_c + s_0) \cdot (d - z) = 0, \tag{13}$$

which yields

$$q_p(z) = \frac{s_c}{s_0} = -\frac{z}{d}. \tag{14}$$

Looking at Eq.(2), one can see that a point charge moving between two planes creates a constant current between the collector and ground during the flight time $\tau$.

Substitution of Eq.(14) into Eq.(11) results in

$$A_0 = -\tfrac{1}{2},\ A_1 = -\tfrac{1}{3}. \tag{15}$$

The final result calculated with Eq.(12) is the rms width of a rectangular pulse:

$$\kappa_p = \frac{1}{\sqrt{12}} \approx 0.29. \tag{16}$$

- *Cylindrical geometry*

If an FFC can be approximated by two concentric infinite cylinders with the beamlet moving radially, the logic described above is still applicable with replacement of the charged plane by a thin charged cylinder. The corresponding function representing the partial charge at the collector $q_{cyl}(z)$ is calculated using the same condition of equal potential of electrodes and yields

$$q_{cyl}(z) = -\frac{\ln\left(\dfrac{R_c + d}{R_c + d - z}\right)}{\ln\left(\dfrac{R_c + d}{Rc}\right)}, \tag{17}$$

where $R_c$ is the radius of the collector, assumed here to be the inner cylinder. Introducing dimensionless parameters

$$\alpha \equiv \frac{d}{R_c}, \quad L \equiv \ln(1+\alpha) \qquad (18)$$

to make expressions more compact, calculations with Eq.(11) give

$$A_0 = -\frac{\alpha - L}{\alpha L}, \quad A_1 = \frac{-3\alpha^2 + 4\alpha L - 2\alpha + 2L}{4\alpha^2 L}, \qquad (19)$$

and the expression for the geometrical coefficient is

$$\kappa_{cyl} = \sqrt{\frac{2L - 2\alpha + \alpha L}{2\alpha L^2}} \quad . \qquad (20)$$

In the limit of $\alpha \to 0$ representing the transition to a plane case, $d \ll R_c$, Eq.(20) gives the same answer as Eq.(16), $\kappa_p = \frac{1}{\sqrt{12}}$.

While Eq.(20) was derived for the practically interesting case with the inner cylinder as a collector, the formula is applicable for the inverted geometry (the outer cylinder as a collector) as well. Following the same procedure, one arrives again to Eq.(20) with $\alpha = -\frac{d}{R_c}$. Note that the value of the coefficient $\kappa_{cyl}$ for a given radii ratio is the same for both cases.

- *Spherical geometry*

For completeness, the geometrical coefficient for a FFC with two spherical electrodes is

$$\kappa_{sph} = \sqrt{\frac{1+\alpha}{\alpha^2}\left[1 - \frac{1+\alpha}{\alpha^2}L^2\right]} \qquad (21)$$

where notation is the same as for the cylindrical case. Similarly, positive $\alpha$ correspond to the case with the collector to be the inner sphere and negative for the opposite case, as well as reproducing the result for the plane electrodes at $\alpha \to 0$.

## 5. Numerical estimations

Fig.2 compares the geometrical coefficients for three 1D cases (Equations (16), (20), and (21)). The coefficient is maximal for the plane electrodes and decreases for cylindrical and spherical geometries.

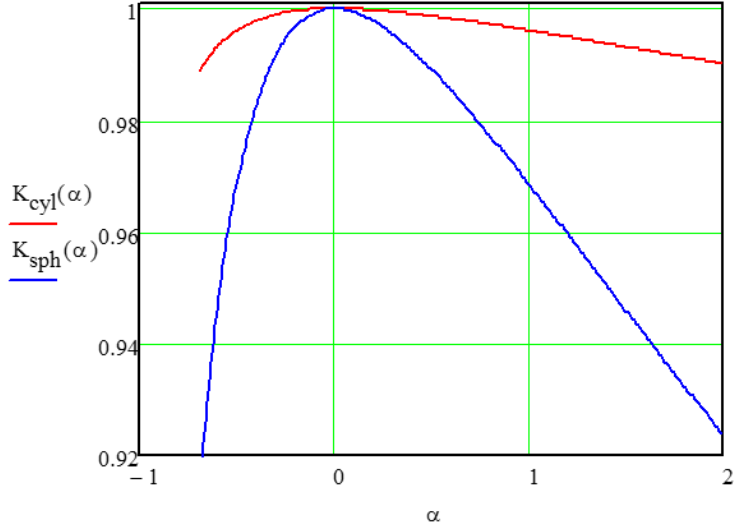

Figure 2. Comparison of three 1D geometries. Geometrical coefficients for the cylindrical (red) and spherical (blue) cases are normalized by the value for the plane electrodes ($1/\sqrt{12}$) and shown as functions of gap-to-collector radius ratio, $\alpha = \pm d/R_c$.

However, for practical applications it is unlikely to have too different radii, and for the radii ratio of <3 (i.e. $-2/3 < \alpha < 2$), the difference between the cylindrical and plane cases is <1%. In part, the FFC shaped as a vacuum coaxial line with 50 Ohm impedance corresponds to $\alpha = 1.3$, and the correction is ~0.6%. Approximations of the model (e.g. effects of the finite orifice hole size) might be noticeably larger. Therefore, in practice using the simple Eq.(16) is sufficient.

Let us consider as an example the case of PIP2IT (formerly PXIE) MEBT [2] with H⁻ ion energy of 2.1 MeV ($v \approx 20$ mm/ns) and a typical rms bunch length of 0.2 ns. For a FFC with the gap $d = 3$ mm, $\tau = 0.15$ ns and $\sigma_{FFC} = \tau/\sqrt{12} \approx 0.04$ ns. Hence, the relative error in measuring the bunch length (Eq.(10)) $\frac{\sigma_c}{\sigma_b} - 1 = \sqrt{1 + \left(\frac{\sigma_{FFC}}{\sigma_b}\right)^2} - 1 \approx 2.3\%$.

## 6. Summary

The rms width of the signal measured by a Fast Faraday Cup is a sum, in quadrature, of the beam rms size and the rms width of the FFC response to a point charge flying through its gap. For practical purposes, the latter can be estimated as $0.3 \cdot \tau$, where $\tau$ is the flight time in the FFC.